\documentclass{PoS}

\usepackage{amsmath}
\usepackage{amssymb}
\usepackage{mathtools}
\usepackage{tikz}
\usetikzlibrary{calc,positioning}
\allowdisplaybreaks

\title{Diagrammatic Coaction of Two-Loop Feynman Integrals}

\ShortTitle{Two-Loop Coaction}

\author{Samuel Abreu\\
        Center for Cosmology, Particle Physics and Phenomenology (CP3), Universit\'e Catholique de
        Louvain, 1348 Louvain-La-Neuve, Belgium\\
        E-mail: \email{samuel.abreu@uclouvain.be}}

\author{Ruth Britto\\
        School of Mathematics and Hamilton Mathematics Institute, Trinity College, Dublin 2, Ireland;\\
        School of Mathematics, Trinity College, Dublin 2, Ireland;\\
        Institut de Physique Th\'eorique, Universit\'e Paris Saclay, CEA, CNRS, F-91191 Gif-sur-Yvette
        cedex, France\\
        E-mail: \email{britto@maths.tcd.ie}}

\author{Claude Duhr\\
        Theoretical Physics Department, CERN, Geneva, Switzerland\\
        E-mail: \email{claude.duhr@cern.ch}}

\author{Einan Gardi\\
        Higgs Centre for Theoretical Physics, School of Physics and Astronomy, The University of
        Edinburgh, Edinburgh EH9 3FD, Scotland, UK\\
        E-mail: \email{einan.gardi@ed.ac.uk}}

\author{\speaker{James Matthew}
	\\
	Higgs Centre for Theoretical Physics, School of Physics and Astronomy, The University of
	Edinburgh, Edinburgh EH9 3FD, Scotland, UK\\
	E-mail: \email{james.matthew@ed.ac.uk}}

\abstract{It is known that one-loop Feynman integrals possess an algebraic structure encoding some of their analytic properties called the coaction, which can be written in terms of Feynman integrals and their cuts. This diagrammatic coaction, and the coaction on other classes of integrals such as hypergeometric functions, may be expressed using suitable bases of differential forms and integration contours. This provides a useful framework for computing coactions of Feynman integrals expressed using the hypergeometric functions. We will discuss examples where this technique has been used in the calculation of two-loop diagrammatic coactions.}

\FullConference{14th International Symposium on Radiative Corrections (RADCOR2019)\\ 
9-13 September 2019\\
		Palais des Papes, Avignon, France}

\begin{document}

\section{Introduction}
The class of iterated integrals known as multiple polylogarithms appears when evaluating Laurent series of certain Feynman integrals in dimensional regularisation. In particular, they are sufficient to express any one-loop Feynman integral. They possess an algebraic structure called the coaction which allows, for instance, simple derivations of functional relations among the polylogarithms \cite{Goncharov:2001iea,Brown:2011ik,Duhr:2012fh}.

Given a Feynman integral expressed using polylogarithms, we may ask if its coaction has any special properties. In fact, at one loop it is understood that the coaction may be expressed in a remarkably simple diagrammatic form featuring Feynman integrals and their cuts \cite{MainCoaction}. More recently, the coactions of various hypergeometric functions have been expressed in a similar form in \cite{Abreu:2019wzk} and explored from a different perspective in \cite{Brown:2019jng}.

Currently, it remains unknown how the diagrammatic coaction generalises beyond one-loop integrals. There is, for instance, the challenge of dealing with more general classes of iterated integral such as elliptic functions. We will not attempt to explore cases which cannot be described by polylogarithms, but even without this complication there are interesting new features in the diagrammatic coaction beyond one loop.

We begin in section \ref{sec2} with an overview of previously published results concerning the coaction of one-loop integrals and hypergeometric functions, before exploring in section \ref{sec3} cuts of two-loop integrals which we will need to provide examples of the two-loop coaction in section \ref{sec4}.
\section{Background}\label{sec2}
\subsection{Multiple Polylogarithms}
The multiple polylogarithms are defined recursively by
\begin{align}
G(a_1,\ldots,a_n;z)=&\int_0^z\frac{dt}{t-a_1}G(a_2,\ldots,a_n;t)\\
\nonumber G(;z)=&1
\end{align}
and possess a coaction \cite{Goncharov:2001iea,Brown:2011ik,Duhr:2012fh} which we will write here in the form
\begin{align}\label{DeltaG}
\Delta G(\vec{a};z)=\sum_{\vec{b}\subseteq\vec{a}}G(\vec{b};z)\otimes G_{\vec{b}}(\vec{a};z),
\end{align}
where the second entries of the coaction $G_{\vec{b}}(\vec{a};z)$ are formed by replacing the integration contour of $G(\vec{a};z)$ with a contour that encircles the poles in $\vec{b}$. For instance
\begin{align}
G_{a_1}(a_1,a_2;z)=\text{Res}_{u=a_1}\int_0^z\frac{du}{u-a_1}\int_0^u\frac{dv}{v-a_2}=\int_0^{a_1}\frac{dv}{v-a_2}=G(a_2;a_1),
\end{align}
where the operator $\text{Res}_{u=a_1}$ replaces the $u$ integration with the residue of the integrand at $u=a_1$. Computing the other terms similarly, we find:
\begin{align}
&\Delta G(a_1,a_2;z)\\
\nonumber=&1\otimes G(a_1,a_2;z)+G(a_1;z)\otimes G(a_2;a_1)+G(a_2;z)\otimes[G(a_1;z)-G(a_1;a_2)]+G(a_1,a_2;z)\otimes 1.
\end{align}
The objects $G_{\vec{b}}(\vec{a};z)$ are described in greater detail in \cite{MainCoaction}.

We note that the coaction of (\ref{DeltaG}) takes the form
\begin{align}\label{mformula}
\Delta\int_{\gamma}\omega=\sum_{i,j}c_{i,j}\int_{\gamma}\omega_i\otimes\int_{\gamma_j}\omega,
\end{align}
with coefficients $c_{i,j}$ determined from the chosen bases of forms and contours $\{\omega_i\}$ and $\{\gamma_j\}$. This form was introduced in \cite{MainCoaction} and the calculation of the $c_{i,j}$ given an alternative description using intersection theory in \cite{Abreu:2019wzk}.
\subsection{One-Loop Graphs}
To each one-loop graph we will associate a unique integral:
\begin{align}
e^{\gamma_E\epsilon}\int\frac{d^Dk}{i\pi^{D/2}}\prod_{i=1}^n\frac{1}{(k+\sum_{j=1}^{i-1}p_j)^2-m_i^2}\qquad\quad D=2\bigg\lceil\frac{n}{2}\bigg\rceil-2\epsilon.
\end{align}
This set of integrals forms a basis for all one-loop integrals, and with our choice of the dimension the integrals are pure after appropriate normalisation. If we denote by $J_E$ the normalised integral associated with a one-loop graph that has a set of edges $E$, then it was argued \cite{MainCoaction} that the coaction of these integrals is
\begin{align}\label{OneLoopDelta}
\Delta J_E=&\sum_{\emptyset\subsetneq X\subseteq E}\left(J_X+a_X\sum_{e\in X}J_{X\backslash e}\right)\otimes\mathcal{C}_XJ_E\\
\nonumber a_X=&\left\{
\begin{array}{cc}
\frac{1}{2}&|X|\text{ even}\\
0&|X|\text{ odd}
\end{array}
\right.,
\end{align}
where $\mathcal{C}_XJ_E$ denotes the cut of the integral $J_E$ on the propagators $X$, i.e. the integral with contour modified to encircle the poles where the propagators in $X$ vanish. We call the terms proportional to $J_{X\backslash e}$ deformation terms.

This result takes the form of (\ref{mformula}), but now each entry of the coaction is a function whose Laurent series in $\epsilon$ has coefficients which are polylogarithms.
\subsection{Hypergeometric Functions}\label{HypCop}
A similar construction has been applied \cite{Abreu:2019wzk} to the integral representations of a number of common hypergeometric functions which expand to polylogarithms: we can find bases of $\{\omega_i\}$ and $\{\gamma_i\}$ and then write the coaction in the form (\ref{mformula}). For example, the Gauss ${}_2F_1$ is the simplest hypergeometric function, and has integral representation
\begin{align}
\frac{\Gamma(a)\Gamma(c-a)}{\Gamma(c)}{}_2F_1(a,b;c;z)=\int_0^1du\,u^{a-1}(1-u)^{c-a-1}(1-uz)^{-b}.
\end{align}
When each of the exponents of the integrand take the form $m+n\epsilon$ for $m,n\in\mathbb{Z}$ we have $\omega=du\, u^{m+a\epsilon}(1-u)^{n+b\epsilon}(1-uz)^{p+c\epsilon}$, $\gamma=[0,1]$ and we can make the choice
\begin{align}
\omega_1&=du\,u^{a\epsilon}(1-u)^{-1+b\epsilon}(1-uz)^{c\epsilon}\\
\nonumber\omega_2&=du\,u^{a\epsilon}(1-u)^{b\epsilon}(1-uz)^{-1+c\epsilon}\\
\nonumber\gamma_1&=b\epsilon[0,1]\\
\nonumber\gamma_2&=c\epsilon z[0,1/z]
\end{align}
to produce the result $c_{i,j}=\delta_{i,j}$. The coaction is then given by
\begin{align}\label{2f1coaction}
&\Delta\left[{}_2F_1\left(m+a\epsilon,n+b\epsilon;p+c\epsilon;z\right)\right]\\
\nonumber=&{}_2F_1\left(1+a\epsilon,b\epsilon;1+c\epsilon;z\right)\otimes {}_2F_1\left(m+a\epsilon,n+b\epsilon;p+c\epsilon;z\right)\\
\nonumber&-\frac{b\epsilon}{1+c\epsilon}{}_2F_1\left(1+a\epsilon,1+b\epsilon;2+c\epsilon;z\right)\\
\nonumber&\otimes\left\{ z^{1-m-a\epsilon}\frac{\Gamma(1-n-b\epsilon)\Gamma(p+c\epsilon)}{\Gamma(1+m-n+(a-b)\epsilon)\Gamma(p-m+(c-a)\epsilon)}\right.\\
\nonumber&\left.\hspace{12pt}{}_2F_1\left(m+a\epsilon,1+m-p+(a-c)\epsilon;1+m-n+(a-b)\epsilon;\frac{1}{z}\right)\right\}.
\end{align}
As Feynman integrals can be expressed using generalised hypergeometric functions \cite{Kalmykov:2009tw}, knowing their coactions in a closed form such as (\ref{2f1coaction}) enables us to determine the coaction of the Feynman integrals without the need for expansion in $\epsilon$. The construction of the coaction on hypergeometric functions is discussed in greater detail in this same volume of proceedings by Ruth Britto.
\section{Cuts of Two-Loop Feynman Integrals}\label{sec3}
\begin{figure}
	\begin{center}
		\begin{tikzpicture}[baseline={([yshift=-.5ex]current bounding box.center)}]
		\coordinate (G1) at (0,0);
		\coordinate (H1) at (2/2,0);
		\coordinate (H2) at (2/2,2/2);
		\coordinate (H3) at (2/2,-2/2);
		\coordinate (G2) at (2,0);
		\coordinate (G3) at (-2/3,0);
		\coordinate (G4) at (8/3,0);
		\coordinate (L1) at (2/2,2/4);
		\coordinate (L2) at (2/2,0);
		\coordinate (L3) at (2/2,-2/3);
		\node at (L1) [above=0 of L1]{$1$};
		\node at (L1) [above=0 of L2]{$2$};
		\node at (L1) [above=1mm of L3]{$3$};
		\draw (G1) [line width=0.75 mm] -- (G2);
		\draw (G1) [line width=0.75 mm] -- (G3);
		\draw (G2) [line width=0.75 mm]-- (G4);
		\draw (G1) to[out=60,in=120] (G2);
		\draw (G1) to[out=-60,in=-120] (G2);
		\end{tikzpicture}
	\end{center}
	\caption{Sunset with one internal mass}\label{fig1}
\end{figure}
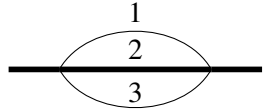
In order to generalise the one-loop coaction (\ref{OneLoopDelta}) to two-loop examples we will need to compute cuts beyond one loop. Take for instance a sunset graph with a single internal mass, shown in figure \ref{fig1} with bold lines denoting non-vanishing internal masses or non-null external momenta. It is known that this topology possesses two master integrals and we will define
\begin{align}
I(m^2,p^2,\nu_1,\nu_2,\nu_3,\nu_4,\nu_5,D)=&\frac{e^{2\gamma_E\epsilon}}{(i\pi^{D/2})^2}\int d^Dk\int d^Dl\frac{[(k+l)^2]^{\nu_4}[(k+p)^2]^{\nu_5}}{[k^2]^{\nu_1} [l^2]^{\nu_2}[(k+l+p)^2-m^2]^{\nu_3}}
\end{align}
and then choose as a basis $I^{(1)}$ and $I^{(2)}$ given by:
\begin{align}
I^{(1)}=&\left(p^2-m^2\right)I(m^2,p^2,1,1,1,0,0,2-2\epsilon)\\
\nonumber=&\left(1-\frac{p^2}{m^2}\right)e^{2\gamma_E\epsilon}(m^2)^{-2\epsilon}\frac{\Gamma^2(-\epsilon)\Gamma(1+2\epsilon)\Gamma(1+\epsilon)}{\Gamma(1-\epsilon)}{}_2F_1\left(1+2\epsilon,1+\epsilon;1-\epsilon;\frac{p^2}{m^2}\right)\\
\nonumber I^{(2)}=&-I(m^2,p^2,1,1,1,1,0,2-2\epsilon)\\
\nonumber=&e^{2\gamma_E\epsilon}(m^2)^{-2\epsilon}\frac{\Gamma^2(-\epsilon)\Gamma(1+2\epsilon)\Gamma(1+\epsilon)}{\Gamma(1-\epsilon)}{}_2F_1\left(2\epsilon,\epsilon;1-\epsilon;\frac{p^2}{m^2}\right).
\end{align}
Applying (\ref{2f1coaction}) and manipulating the functions which appear in the coaction reveals that
\begin{align}\label{DeltaI1Raw}
\Delta I^{(1)}=&I^{(1)}\otimes e^{2\gamma_E}\frac{\Gamma^2(1-\epsilon)}{\Gamma(1-4\epsilon)}(p^2)^{2\epsilon}\left(m^2-p^2\right)^{-4\epsilon}{}_2F_1\left(-2\epsilon,-\epsilon;-4\epsilon;1-\frac{m^2}{p^2}\right)\\
\nonumber&+I^{(2)}\otimes \left[e^{2\gamma_E\epsilon}\frac{\Gamma(1-\epsilon)\Gamma(1+\epsilon)}{\Gamma(1-2\epsilon)}(m^2-p^2)^{-2\epsilon}{}_2F_1\left(-2\epsilon,1+2\epsilon;1-\epsilon;\frac{p^2}{p^2-m^2}\right)\right.\\
\nonumber&\left.-e^{2\gamma_E}\frac{\Gamma^2(1-\epsilon)}{\Gamma(1-4\epsilon)}(p^2)^{2\epsilon}\left(m^2-p^2\right)^{-4\epsilon}{}_2F_1\left(-2\epsilon,-\epsilon;-4\epsilon;1-\frac{m^2}{p^2}\right)\right].
\end{align}
As each term in this coaction has one of the integrals $I^{(1)}$ or $I^{(2)}$ in the first entry we would like to, by analogy to the one-loop case, interpret each of the second entries as a maximal cut of the integral $I^{(1)}$. This is because, at one loop, the maximal cut is paired with the graph that has no propagators contracted (see (\ref{OneLoopDelta})).

Let us compute the maximal cut of $I^{(1)}$ which corresponds to its discontinuity with respect to $p^2$. We can compute this by parametrising the momenta and taking residues. We write the cut as
\begin{align}
\mathcal{C}_3\int d^Dl\frac{1}{l^2}\mathcal{C}_{1,2}\int d^Dk\frac{1}{[(k+l+p)^2-m^2][k^2]},
\end{align}
where $\mathcal{C}_{ijk\ldots}$ denotes the operation of cutting propagators $i$, $j$, $k$,$\ldots$. Then we can perform the cut calculation for the inner loop in a frame where $l+p$ has a single non-vanishing component \cite{MainCuts}:
\begin{align}
l+p=&\sqrt{(l+p)^2}(1,\vec{0}),\\
k=&k_0(1,\beta\vec{1}_{D-1}),\\
d^Dk=&\frac{2\pi^{\frac{D-1}{2}}}{\Gamma\left(\frac{D-1}{2}\right)}dk_0k_0^{D-1}d\beta\,\beta^{D-2}.
\end{align}
The cut of the first loop is
\begin{align}
&\mathcal{C}_{1,2}\int d^Dk\frac{1}{[(k+l+p)^2-m^2][k^2]}\\
\nonumber=&\frac{2\pi^{\frac{D-1}{2}}}{\Gamma\left(\frac{D-1}{2}\right)}\text{Res}_{k_0=\frac{m^2-(l+p)^2}{2\sqrt{(l+p)^2}}}\text{Res}_{\beta=1}\int dk_0k_0^{D-1}\int d\beta\,\beta^{D-2}\\
\nonumber&\frac{1}{[k_0^2(1-\beta^2)+(l+p)^2+2\sqrt{(l+p)^2}k_0-m^2][k_0^2(1-\beta^2)]}\\
\nonumber=&-\frac{\pi^{\frac{D-1}{2}}}{\Gamma\left(\frac{D-1}{2}\right)}\frac{1}{2\sqrt{(l+p)^2}}\left[\frac{m^2-(l+p)^2}{2\sqrt{(l+p)^2}}\right]^{D-3}.
\end{align}
Then we handle the outer loop similarly, inserting a $\theta((l+p)^2-m^2)$ to account for the absence of a discontinuity in the inner loop when the momentum flowing into it is below threshold:
\begin{align}\label{SunsetMaxCut}
&\mathcal{C}_3\int d^Dl\frac{1}{l^2}\mathcal{C}_{1,2}\int d^Dk\frac{1}{[(k+l+p)^2-m^2][k^2]}\\
\nonumber=&-\frac{2\pi^{D-1}}{\Gamma^2\left(\frac{D-1}{2}\right)}\text{Res}_{\beta=1}\int dl_0\,l_0^{D-1}\int d\beta\,\beta^{D-2}\frac{1}{l_0^2(1-\beta^2)}\frac{\theta((l+p)^2-m^2)}{2\sqrt{(l+p)^2}}\left[\frac{m^2-(l+p)^2}{2\sqrt{(l+p)^2}}\right]^{D-3}\\
\nonumber=&\frac{1}{2^{D-2}}\frac{\pi^{D-1}}{\Gamma^2\left(\frac{D-1}{2}\right)}\int_{\frac{m^2-p^2}{2\sqrt{p^2}}}^0 dl_0\,l_0^{D-3}[p^2+2\sqrt{p^2}l_0]^{1-D/2}\left[m^2-p^2-2\sqrt{p^2}l_0\right]^{D-3}\\
\nonumber=&\frac{\Gamma^2(D-2)\Gamma^2(D/2)}{\Gamma(2D-4)\Gamma^2(D-1)}\pi^{D-2}(p^2)^{2-D}(m^2-p^2)^{2D-5}{}_2F_1\left(D-2,D/2-1;2D-4;1-\frac{m^2}{p^2}\right).
\end{align}
We note that in the coaction of the ${}_2F_1$, the second entries were formed by modifying the integration contour while keeping the same integrand. We can similarly define a second maximal cut of the sunset by changing the region of momentum space over which we integrate in (\ref{SunsetMaxCut}). The relevant endpoints can be identified in the $l_0$ integrand and the new cut is:
\begin{align}
&\frac{1}{2^{D-2}}\frac{\pi^{D-1}}{\Gamma^2\left(\frac{D-1}{2}\right)}\int_0^{-\frac{\sqrt{p^2}}{2}} dl_0\,l_0^{D-3}[p^2+2\sqrt{p^2}l_0]^{1-D/2}\left[m^2-p^2-2\sqrt{p^2}l_0\right]^{D-3}\\
\nonumber=&\frac{\Gamma(D/2)\Gamma(D-2)\Gamma(2-D/2)}{\Gamma^2(D-1)}\pi^{D-2} (m^2-p^2)^{D-3}{}_2F_1\left(D-2,3-D;D/2;\frac{p^2}{p^2-m^2}\right).
\end{align}
Then after normalising by $(2\pi i)^2$ we have a set of cuts of $I^{(1)}$ given by:
\begin{align}\label{Cuts}
&e^{2\gamma_E\epsilon}\frac{4}{\epsilon}\frac{\Gamma^2(1-\epsilon)}{\Gamma(1-4\epsilon)}(p^2)^{2\epsilon}(m^2-p^2)^{-4\epsilon}{}_2F_1\left(-2\epsilon,-\epsilon;-4\epsilon;1-\frac{m^2}{p^2}\right),\\
\nonumber&e^{2\gamma_E\epsilon}\frac{2}{\epsilon}\frac{\Gamma(1-\epsilon)\Gamma(1+\epsilon)}{\Gamma(1-2\epsilon)}(m^2-p^2)^{-2\epsilon}{}_2F_1\left(-2\epsilon,1+2\epsilon;1-\epsilon;\frac{p^2}{p^2-m^2}\right).
\end{align}

\section{Two-Loop Diagrammatic Coaction}\label{sec4}

Linear combinations of the cuts in (\ref{Cuts}) appear in the second entries of $\Delta I^{(1)}$ in (\ref{DeltaI1Raw}) and we then define cuts $\mathcal{C}^{(1)}I^{(1)}$ and $\mathcal{C}^{(2)}I^{(1)}$ which are exactly these linear combinations. The coaction is then of the form (\ref{mformula}):
\begin{align}\label{Suns1}
\Delta I^{(1)}=I^{(1)}\otimes\mathcal{C}^{(1)}I^{(1)}+I^{(2)}\otimes\mathcal{C}^{(2)}I^{(1)}.
\end{align}
We may compute the cuts of $I^{(2)}$ with the same contours and find the coaction:
\begin{align}\label{Suns2}
\Delta I^{(2)}=I^{(1)}\otimes\mathcal{C}^{(1)}I^{(2)}+I^{(2)}\otimes\mathcal{C}^{(2)}I^{(2)}.
\end{align}

We are free to choose a basis other than $I^{(1)}$, $I^{(2)}$, $\mathcal{C}^{(1)}$ and $\mathcal{C}^{(2)}$. We may change, say, the integrands by applying a rotation matrix $\mathcal{M}$ to the vector of forms $(\omega_1,\omega_2)$ but still retain the same form of the coaction if we transform the contours with $(\mathcal{M}^{-1})^T$.

Diagrammatically, we can write this coaction as:
\begin{align*}
\Delta\left[\begin{tikzpicture}[baseline={([yshift=-.5ex]current bounding box.center)}]
\coordinate (G1) at (0,0);
\coordinate (H1) at (1/2,0);
\coordinate (H2) at (1/2,1/2);
\coordinate (H3) at (1/2,-1/2);
\coordinate (G2) at (1,0);
\coordinate (G3) at (-1/3,0);
\coordinate (G4) at (4/3,0);
\node at (G2) [above=1mm of G2]{\color{red}\small$(1)$};
\node at (G2) [below=1mm of G2]{\vphantom{(1)}};
\draw (G1) [line width=0.75 mm] -- (G2);
\draw (G1) [line width=0.75 mm] -- (G3);
\draw (G2) [line width=0.75 mm]-- (G4);
\draw (G1) to[out=60,in=120] (G2);
\draw (G1) to[out=-60,in=-120] (G2);
\end{tikzpicture}\right]=&\begin{tikzpicture}[baseline={([yshift=-.5ex]current bounding box.center)}]
\coordinate (G1) at (0,0);
\coordinate (H1) at (1/2,0);
\coordinate (H2) at (1/2,1/2);
\coordinate (H3) at (1/2,-1/2);
\coordinate (G2) at (1,0);
\coordinate (G3) at (-1/3,0);
\coordinate (G4) at (4/3,0);
\node at (G2) [above=1mm of G2]{\color{red}\small$(1)$};
\node at (G2) [below=1mm of G2]{\vphantom{(1)}};
\draw (G1) [line width=0.75 mm] -- (G2);
\draw (G1) [line width=0.75 mm] -- (G3);
\draw (G2) [line width=0.75 mm]-- (G4);
\draw (G1) to[out=60,in=120] (G2);
\draw (G1) to[out=-60,in=-120] (G2);
\end{tikzpicture}\otimes\begin{tikzpicture}[baseline={([yshift=-.5ex]current bounding box.center)}]
\coordinate (G1) at (0,0);
\coordinate (H1) at (1/2,0);
\coordinate (H2) at (1/2,1/2);
\coordinate (H3) at (1/2,-1/2);
\coordinate (G2) at (1,0);
\coordinate (G3) at (-1/3,0);
\coordinate (G4) at (4/3,0);
\node at (G2) [above=1mm of G2]{\color{red}\small$(1)$};
\node at (G2) [below=1mm of G2]{\vphantom{(1)}};
\draw (G1) [line width=0.75 mm] -- (G2);
\draw (G1) [line width=0.75 mm] -- (G3);
\draw (G2) [line width=0.75 mm]-- (G4);
\draw (G1) to[out=60,in=120] (G2);
\draw (G1) to[out=-60,in=-120] (G2);
\draw (H3) [dashed,color=red,line width=0.5 mm] --(H2);
\end{tikzpicture}+\begin{tikzpicture}[baseline={([yshift=-.5ex]current bounding box.center)}]
\coordinate (G1) at (0,0);
\coordinate (H1) at (1/2,0);
\coordinate (H2) at (1/2,1/2);
\coordinate (H3) at (1/2,-1/2);
\coordinate (G2) at (1,0);
\coordinate (G3) at (-1/3,0);
\coordinate (G4) at (4/3,0);
\node at (G2) [above=1mm of G2]{\color{blue}\small$(2)$};
\node at (G2) [below=1mm of G2]{\vphantom{(1)}};
\draw (G1) -- (G2);
\draw (G1) [line width=0.75 mm] -- (G2);
\draw (G1) [line width=0.75 mm] -- (G3);
\draw (G2) [line width=0.75 mm]-- (G4);
\draw (G1) to[out=60,in=120] (G2);
\draw (G1) to[out=-60,in=-120] (G2);
\end{tikzpicture}\otimes\begin{tikzpicture}[baseline={([yshift=-.5ex]current bounding box.center)}]
\coordinate (G1) at (0,0);
\coordinate (H1) at (1/2,0);
\coordinate (H2) at (1/2,1/2);
\coordinate (H3) at (1/2,-1/2);
\coordinate (G2) at (1,0);
\coordinate (G3) at (-1/3,0);
\coordinate (G4) at (4/3,0);
\node at (G2) [above=1mm of G2]{\color{red}\small$(1)$};
\node at (G2) [below=1mm of G2]{\vphantom{(1)}};
\draw (G1) [line width=0.75 mm] -- (G2);
\draw (G1) [line width=0.75 mm] -- (G3);
\draw (G2) [line width=0.75 mm]-- (G4);
\draw (G1) to[out=60,in=120] (G2);
\draw (G1) to[out=-60,in=-120] (G2);
\draw (H3) [dashed,color=blue,line width=0.5 mm] --(H2);
\end{tikzpicture}\\
\nonumber\Delta\left[\begin{tikzpicture}[baseline={([yshift=-.5ex]current bounding box.center)}]
\coordinate (G1) at (0,0);
\coordinate (H1) at (1/2,0);
\coordinate (H2) at (1/2,1/2);
\coordinate (H3) at (1/2,-1/2);
\coordinate (G2) at (1,0);
\coordinate (G3) at (-1/3,0);
\coordinate (G4) at (4/3,0);
\node at (G2) [above=1mm of G2]{\color{blue}\small$(2)$};
\node at (G2) [below=1mm of G2]{\vphantom{(1)}};
\draw (G1) [line width=0.75 mm] -- (G2);
\draw (G1) [line width=0.75 mm] -- (G3);
\draw (G2) [line width=0.75 mm]-- (G4);
\draw (G1) to[out=60,in=120] (G2);
\draw (G1) to[out=-60,in=-120] (G2);
\end{tikzpicture}\right]=&\begin{tikzpicture}[baseline={([yshift=-.5ex]current bounding box.center)}]
\coordinate (G1) at (0,0);
\coordinate (H1) at (1/2,0);
\coordinate (H2) at (1/2,1/2);
\coordinate (H3) at (1/2,-1/2);
\coordinate (G2) at (1,0);
\coordinate (G3) at (-1/3,0);
\coordinate (G4) at (4/3,0);
\node at (G2) [above=1mm of G2]{\color{red}\small$(1)$};
\node at (G2) [below=1mm of G2]{\vphantom{(1)}};
\draw (G1) [line width=0.75 mm] -- (G2);
\draw (G1) [line width=0.75 mm] -- (G3);
\draw (G2) [line width=0.75 mm]-- (G4);
\draw (G1) to[out=60,in=120] (G2);
\draw (G1) to[out=-60,in=-120] (G2);
\end{tikzpicture}\otimes\begin{tikzpicture}[baseline={([yshift=-.5ex]current bounding box.center)}]
\coordinate (G1) at (0,0);
\coordinate (H1) at (1/2,0);
\coordinate (H2) at (1/2,1/2);
\coordinate (H3) at (1/2,-1/2);
\coordinate (G2) at (1,0);
\coordinate (G3) at (-1/3,0);
\coordinate (G4) at (4/3,0);
\node at (G2) [above=1mm of G2]{\color{blue}\small$(2)$};
\node at (G2) [below=1mm of G2]{\vphantom{(1)}};
\draw (G1) [line width=0.75 mm] -- (G2);
\draw (G1) [line width=0.75 mm] -- (G3);
\draw (G2) [line width=0.75 mm]-- (G4);
\draw (G1) to[out=60,in=120] (G2);
\draw (G1) to[out=-60,in=-120] (G2);
\draw (H3) [dashed,color=red,line width=0.5 mm] --(H2);
\end{tikzpicture}+\begin{tikzpicture}[baseline={([yshift=-.5ex]current bounding box.center)}]
\coordinate (G1) at (0,0);
\coordinate (H1) at (1/2,0);
\coordinate (H2) at (1/2,1/2);
\coordinate (H3) at (1/2,-1/2);
\coordinate (G2) at (1,0);
\coordinate (G3) at (-1/3,0);
\coordinate (G4) at (4/3,0);
\node at (G2) [above=1mm of G2]{\color{blue}\small$(2)$};
\node at (G2) [below=1mm of G2]{\vphantom{(1)}};
\draw (G1) [line width=0.75 mm] -- (G2);
\draw (G1) [line width=0.75 mm] -- (G3);
\draw (G2) [line width=0.75 mm]-- (G4);
\draw (G1) to[out=60,in=120] (G2);
\draw (G1) to[out=-60,in=-120] (G2);
\end{tikzpicture}\otimes\begin{tikzpicture}[baseline={([yshift=-.5ex]current bounding box.center)}]
\coordinate (G1) at (0,0);
\coordinate (H1) at (1/2,0);
\coordinate (H2) at (1/2,1/2);
\coordinate (H3) at (1/2,-1/2);
\coordinate (G2) at (1,0);
\coordinate (G3) at (-1/3,0);
\coordinate (G4) at (4/3,0);
\node at (G2) [above=1mm of G2]{\color{blue}\small$(2)$};
\node at (G2) [below=1mm of G2]{\vphantom{(1)}};
\draw (G1) [line width=0.75 mm] -- (G2);
\draw (G1) [line width=0.75 mm] -- (G3);
\draw (G2) [line width=0.75 mm]-- (G4);
\draw (G1) to[out=60,in=120] (G2);
\draw (G1) to[out=-60,in=-120] (G2);
\draw (H3) [dashed,color=blue,line width=0.5 mm] --(H2);
\end{tikzpicture}
\end{align*}
where we indicate the first elements in the basis of integrands and cut contours with the colour red and the second elements with blue.

In this example we see a feature that did not occur in one-loop cases: there is a sum of terms where the master integrals in the first entry correspond to the same graph and the cuts in the second entry correspond to placing the same collection of propagators on shell. The pairing, observed for one-loop integrals in (\ref{OneLoopDelta}), between integrals $J_X$ in the first entry and cuts $\mathcal{C}_XJ_E$ in the second entry is preserved in this case.

Let us now consider a number of other examples to see how these features generalise. In each case, we find a suitable basis of integrals, evaluate them as hypergeometric functions and use the technique of \ref{HypCop} to find the coaction. We frequently have to use well-known identities on these hypergeometric functions in order to relate terms in the coaction to Feynman integrals and their cuts \cite{BaileyHyper,HYPERDIRE}.

For the sunset graph with two masses, we find a similar structure as the one-mass case, but with the addition of deformation terms. There is the same invariance under rotating to a different basis as in the one-mass case, and we find that for any such choice that is normalised suitably we get the coaction
\begin{align*}
\Delta\left[

\end{align*}
where the third integral and cut are now denoted by the colour green, with red and blue denoting the first two as before.

The deformation terms here have a coefficient of $1$, which differs from the $\frac{1}{2}$ found at one loop. We also note that they are formed by contracting a propagator in a loop which has an even number of propagators as in (\ref{OneLoopDelta}).

We can also explore two-loop cases where the number of propagators is larger. For instance, the diagrammatic coaction of a triangle graph with a double edge are shown for various kinematic configurations below.
\begin{align*}
\Delta\left[

\end{align*}

In each of the cases we have presented above, it can be demonstrated that limits in the kinematic variables or the masses commute with the coaction, a behaviour that was also present at one loop.

Lastly we give a pair of examples where there are five propagators. In each case we find the same structure as before continues to hold.
\begin{align*}
&\Delta\left[%
\end{align*}
\section{Summary and Outlook}

We have outlined the coaction of multiple polylogarithms and how it applies to the cases of one-loop graphs and certain hypergeometric functions.

We have also described the coactions of a number of two-loop Feynman integrals. These coactions possess a structure that is similar to the one-loop case: the pairing between contracted and cut graphs is preserved, with the possibility of deformation terms. But there are new features also, such as the sum over different master integrals for the same graph and their corresponding cuts.

It will be desirable to explore the circumstances under which we obtain deformation terms and what their coefficients are. At one loop there are homology relations between the cut contours which explain the origin of the deformation terms, but the generalisation of these results to two loops remains to be established.

We also leave open the case of Feynman integrals that are not expressible using only MPLs. It is unclear if a diagrammatic coaction exists in these cases, and if such a coaction could take the same form as the two-loop examples we have presented despite the difference in the underlying functions.

\acknowledgments
This work is supported by the ``Fonds National de la Recherche Scientifique'' (FNRS), Belgium (SA),  by the ERC Consolidator Grant 647356 ``CutLoops'' (RB), the ERC Starting Grant 637019 ``MathAm'' (CD), and the STFC Consolidated Grant ``Particle Physics at the Higgs Centre'' (EG, JM).


\bibliographystyle{JHEP}
\bibliography{BibMaster}


\end{document}